\documentclass[particles,review,accept,pdftex,moreauthors]{Definitions/mdpi} 
\usepackage{amssymb,slashed,color,graphicx,bm,mathrsfs,color}
\newcommand{\be}{\begin{equation}}
\newcommand{\ee}{\end{equation}}
\newcommand{\bea}{\begin{eqnarray}}
\newcommand{\eea}{\end{eqnarray}}

\newcommand{\veca}{{\bm \alpha}}
\newcommand{\vecq}{\bm q}
\newcommand{\gammavec}{{\bm \gamma}}
\newcommand{\veck}{{\bm k}}
\newcommand{\ep}{{\epsilon}}

\def\aap{Astron. Astrophys.}%
\def\mnras{Mon. Not. R. Astron. Soc.}%
\def\prc{Phys.~Rev.~C~}%
\def\prd{Phys.~Rev.~D~}%
\def\prl{Phys.~Rev.~Lett.}%
\def\nphysa{Nucl.~Phys.~A}%
\def\physrep{Phys.~Rep.~}%
\usepackage[normalem]{ulem}  

\firstpage{1} 
\pubvolume{1}
\issuenum{1}
\articlenumber{0}
\pubyear{2023}
\copyrightyear{2023}
\externaleditor{Academic Editor: Firstname Lastname}
\datereceived{25 March 2023} 
\daterevised{9 June 2023 } 
\dateaccepted{11 July 2023 } 
\datepublished{ } 
\hreflink{https://doi.org/} 

\Title{Impact of Multiple Phase Transitions in Dense QCD on Compact Stars}

\TitleCitation{Impact of Multiple Phase Transitions in Dense QCD on Compact Stars}

\Author{Armen Sedrakian $^{1,2}$\orcidA{}}

\AuthorNames{Armen Sedrakian}

\AuthorCitation{Sedrakian, A.}

\address{%
$^{1}$ \quad Frankfurt Institute for Advanced Studies, D-60438
  Frankfurt am Main, Germany; sedrakian@fias.uni-frankfurt.de or armen.sedrakian@uwr.edu.pl
$^{2}$ \quad Institute of Theoretical Physics, University of Wroc\l{}aw,
50-204 Wroc\l{}aw, Poland}

\abstract{ This review covers several recent developments in the
  physics of dense QCD with an emphasis on the impact of multiple
  phase transitions on astrophysical manifestations of compact
  stars. To motivate the multi-phase modeling of dense QCD and
  delineate the perspectives, we start with a discussion of the
  structure of its phase diagram and the arrangement of possible
  color-superconducting and other phases. It is conjectured that
  pair-correlated quark matter in $\beta$-equilibrium is within the
  same universality class as spin-imbalanced cold atoms and the
  isospin asymmetrical nucleonic matter. This then implies the
  emergence of phases with broken space symmetries and tri-critical
  (Lifshitz) points.  The beyond-mean-field structure of the quark
  propagator and its non-trivial implications are discussed in the
  cases of two- and three-flavor quark matter within the Eliashberg
  theory, which takes into account the frequency dependence
  (retardation) of the gap function. We then construct an equation of
  state (EoS) that extends the two-phase EoS of dense quark matter
  within the constant speed of sound parameterization by adding a
  conformal fluid with a speed of sound $c_{\rm conf.}=1/\sqrt{3}$ at
  densities $\ge 10~n_{\rm sat}$, where $n_{\rm sat}$ is the
  saturation density. With this input, we construct static,
  spherically symmetrical compact hybrid stars in the mass--radius
  diagram, recover such features as the twins and triplets, and show
  that the transition to conformal fluid leads to the spiraling-in of
  the tracks in this diagram. Stars on the spirals are classically
  unstable with respect to the radial oscillations but can be
  stabilized if the conversion timescale between quark and nucleonic
  phases at their interface is larger than the oscillation period.
  Finally, we review the impact of a transition from high-temperature
  gapped to low-temperature gapless two-flavor phase on the thermal
  evolution of hybrid stars.  }


\begin{document}

\maketitle

\section{Introduction}
\label{sec:intro}

The astrophysics of compact stars entered the era of multimessenger astronomy in 2017 with the discovery of the neutron star binary merger event GW170817~\cite{LIGO_Virgo2017}. Combined with radio observations of massive pulsars in binaries with white dwarfs~\cite{NANOGrav2019} and X-ray observations of nearby solitary neutron stars~\cite{NICER2021a,NICER2021b}, compact star astrophysics nowadays offers  important insights into their global properties and potentially into the phase structure of dense matter~\cite{Weber2005PrPNP,Alford2008,Fukushima2011RPPh,Anglani2014,Blaschke2018,Baym2018RPPh,Burgio2021PrPNP,Lattimer2023Parti,Sedrakian2023PrPNP}.  Studies of matter at high densities are fundamental to our understanding of the strong force of the Standard Model and underlying concepts such as confinement, spontaneous chiral symmetry breaking, and dynamical mass generation~\cite{Ding2023Parti}. 

This work studies the impact of multiple phase transitions in dense QCD matter on the physics of compact stars. We partially review the relevant physics but also provide a novel discussion of the mass--radius diagram of compact stars
in the case where a conformal fluid is added to the two-phase, constant speed of sound parametrization of the EoS of quark matter~\cite{Alford2017PRL}.  This is motivated by the recent work that showed that even though the high-central-density stars are typically unstable toward radial oscillation (see Ref.~\cite{Bardeen1977}, hereafter BTM), i.e., when $dM/d \rho_c<0$, where $M$ is the star's gravitational mass and  $\rho_c$ is the central density, they can be stabilized if the conversion between nucleonic and quark phases is slow compared to the characteristic period of radial oscillations~\cite{Pereira2018,Curin2021,Goncalves2022,Rau2022}.

To motivate the modeling, Section~\ref{sec:phases} provides a brief
overview of the phase diagram of dense QCD matter as we understand it
from the studies of the thermodynamics of various high-density phases,
such as the color-superconducting
phases~\cite{Alford2008,Fukushima2011RPPh,Anglani2014,Blaschke2018,Baym2018RPPh}
or quarkionic
phases~\cite{McLerran2007,Fukushima2016ApJ,Pisarski2019,McLerran2019PhRvL,Margueron2021PhRvC}.
Utilizing the knowledge gained from the studies of imbalanced cold
atoms~\cite{Strinati2018} and isospin asymmetrical nuclear
matter~\cite{Stein2014PhRvC,Sedrakian2019EPJA}, the possible phase
structure of pair-correlated quark matter is conjectured based on the
universal features of imbalanced pair-correlated fermionic systems.
Section~\ref{sec:GF} discusses the computations of Green's functions
in the two- and three-flavor
phases~\cite{Sedrakian2018PhLB,AlfordWindisch2018} and potential new
effects arising beyond the adiabatic (frequency-independent)
approximation of the gap. In Section~\ref{sec:MR}, the constant
speed-of-sound parameterization of the EoS of quark matter
phases~\cite{Zdunik2013,Han2013PhRvD} is used to explore the
mass--radius ($M$-$R$) diagram of compact stars with deconfinement and
multiple phase transitions. For two-phase transitions, one from
nucleonic to two-flavor quark matter and another from two-flavor to
three-flavor phase of quark matter, we recover the {\it fourth family
  of compact stars, } which is separated from the third family by the
instability region~\cite{Alford2017PRL}. Here we show that a
high-density phase of conformal fluid at densities $\ge 10~n_{\rm
  sat}$, where \mbox{$n_{\rm sat}=0.16$ fm$^{-3}$} is the saturation
density, modifies the classically unstable tracks in the $M$-$R$
diagram compared to the case when such transition does not occur
~\cite{Li2023PhRvD}. Such modification is phenomenologically important
because of the possible stabilization mechanism of radial oscillation
modes of hybrid
stars~\cite{Pereira2018,Curin2021,Goncalves2022,Rau2022}, which is
discussed in Section~\ref{sec:modes}.  In Section~\ref{sec:cooling},
we discuss the cooling of compact stars with quark cores. We then
simulate the thermal evolution of these stars on a time scale on the
order of million years with a focus on the impact of the phase
transition from the gapped to the gapless phase of 2SC matter in the
core of the star.  Our conclusions are given in
Section~\ref{sec:conclusions}.

\section{A Brief Review of the Phase Diagram of Dense QCD}
\label{sec:phases}

\begin{figure}[H]
\includegraphics[width=0.85\hsize,angle=0]{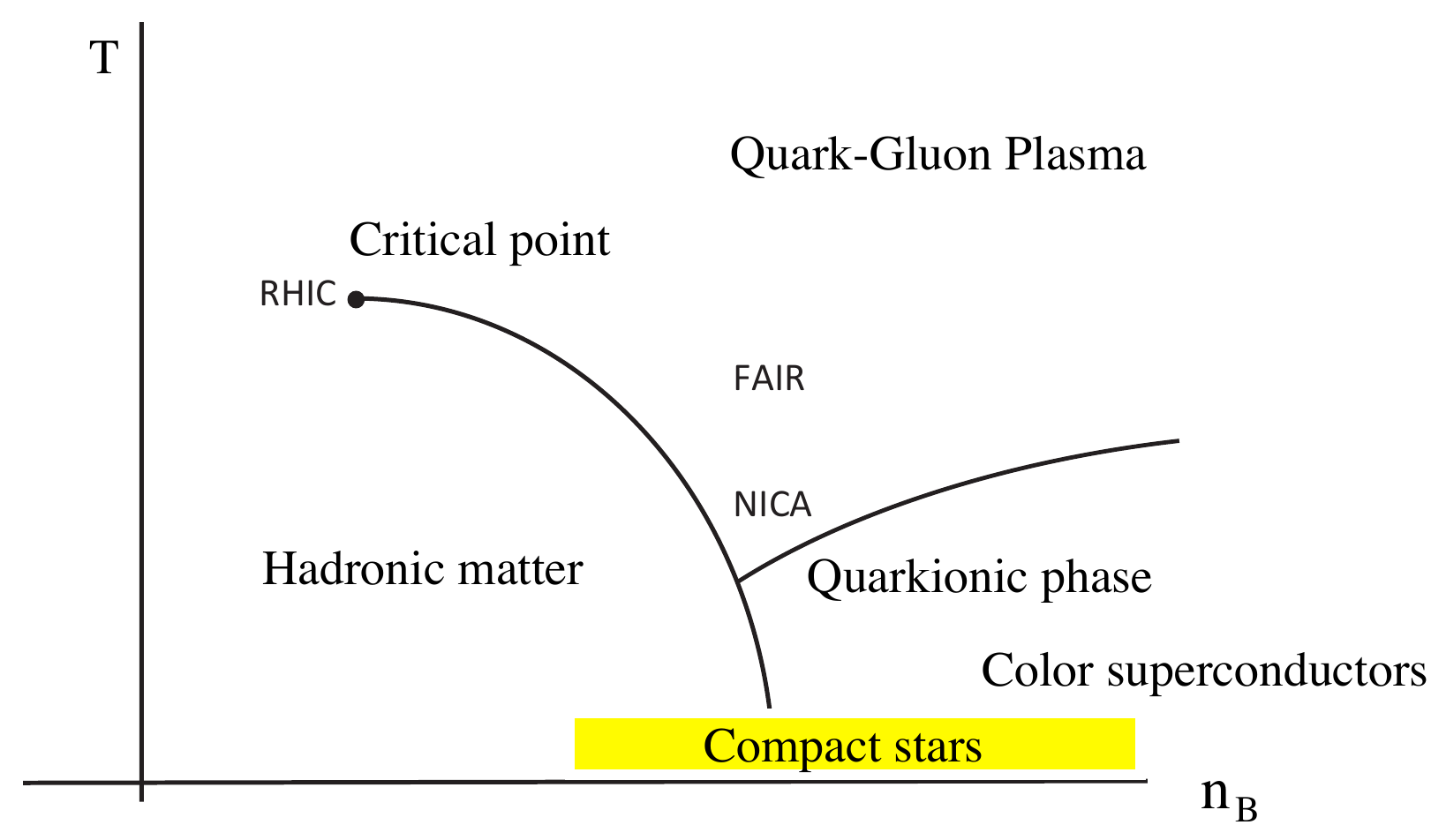}
\caption{Sketch 
 of the phase diagram of strongly interacting matter in the temperature
  and baryonic density plain.  Compact stars cover the low-temperature and high-density regimes of this phase diagram. The parameter ranges covered by the FAIR, NICA, and RHIC facilities are also indicated.}
\label{fig:1}
\end{figure}

Matter in compact stars covers the large number density ($n\ge n_{\rm sat}$), large isospin, and relatively low-temperature ($0\le T\le 100$~MeV) portion of the phase diagram of strongly interacting matter.  The extremely low temperature ($T\le 0.1$) The MeV regime is relevant for mature compact stars, whereas the higher temperature domain is relevant for supernovas and binary neutron star mergers. The complexity of the phase diagram arises due to the multiple order parameters describing (interrelated) phenomena, which include deconfinement phase transition (with the Polyakov loop as the order parameter of the center symmetry), chiral phase transition (and its condensate as the order parameter), the
color-superconducting phases (with the anomalous correlator as the order parameter). Depending on the non-zero value of one or several order parameters, distinct phases may arise: an extensively studied case is color-superconducting phases with various pairing patterns~\cite{Alford2008,Fukushima2011RPPh,Anglani2014,Blaschke2018,Baym2018RPPh}. A more recent suggestion is a confined but chirally symmetric quarkyonic phase at compact star densities~\cite{McLerran2007,Pisarski2019,McLerran2019PhRvL}.  A crude sketch of the phase diagram of strongly interacting matter is shown Figure~\ref{fig:1}, along with the regions that are covered by current and future facilities (RHIC, NICA, and FAIR). The low-density and low-temperature region of the phase diagram contains nuclei embedded into a sea of charge-neutralizing electrons and neutrons at higher densities.  As the density increases, a first-order phase transition to bulk nuclear matter occurs at around $0.5~n_{\rm sat}$. A further increase in density can lead to the deconfinement of nucleons to form quark matter for $n \ge (2-3)\times n_{\rm sat}$.

The transition from nuclear matter to deconfined quark matter could be of the first or second order, or a crossover~\cite{Alford2008,Fukushima2011RPPh,Anglani2014,Blaschke2018,Baym2018RPPh}.  The first-order phase transition leaves a marked imprint on the macroscopic properties of compact stars because the EoS contains a density jump, which may give rise to new stable branches of compact stars (i.e., their third family) separated from nucleonic stars by a region of instability~\cite{Gerlach1968,Glendenning2000,Schertler2000,Alvarez-Castillo2019}. Smooth crossover without changes in the values of the order parameter or the wave function of the three-quark states would be a less dramatic change in the slope of the EoS, best visualized in terms of the  speed of sound~\cite{Baym2018RPPh,Fukushima2016ApJ,Kojo2019AIPC}.
As mentioned above, two sequential first-order phase transitions can lead to the appearance of
a new branch of compact stars---fourth family---separated from the third family by an instability region~\cite{Alford2017PRL,Lijj2020a,Lijj2021,Li2023PhRvD}, assuming the classical stability criterion
$dM/d \rho_c>0$ is valid. In the case of slow phase transition between the nuclear and quark matter phases, the two families are not separated by an instability region; i.e., they form a continuous branch where the regions with $dM/d \rho_c<0$ are stabilized~\cite{Rau2022} (see Section~\ref{sec:modes}).

Deconfined quark matter at low temperature and high density is expected to have characteristic features of degenerate Fermi systems, which are familiar from condensed matter physics. Therefore, emergent phenomena such as (color) superconductivity are expected in channels where gluon exchange is attractive~\cite{Alford2008,Fukushima2011RPPh,Anglani2014}. Various color superconducting phases may arise, depending on the number of flavors and colors involved in the pairing, the ratio of $\Delta/\delta\mu$, where $\Delta$ is the gap in the quasiparticle spectrum, $\delta\mu =(\mu_d-\mu_d)/2$ is the difference in the chemical potentials of down ($d$) and up ($u$) quarks, and the mass of strange quark $m_s$ in the three-flavor quark matter. The two-flavor candidate phases classified according to these parameters are as follows:

\begin{itemize}
\item[(a)]    The ``2SC'' phase 
 (where the abbreviation refers to two-superconducting-colors)~\cite{Alford2008}
  \begin{eqnarray}\label{eq:QCD_2SC} 
\Delta_{2SC} \propto \langle \psi^T(x)C\gamma_5\tau_2\lambda_2\psi(x)\rangle \neq 0 , \quad 0\le \delta\mu <\Delta/\sqrt{2}, 
  \end{eqnarray}
  where $C=i\gamma^2\gamma^0$ is the charge conjugation operator, $\tau_2$ is the second component of the Pauli matrix acting in the SU$(2)_f$ flavor space, and $\lambda_A$ is the antisymmetric Gell--Mann matrix acting in the SU$(3)_c$ color space. The properties of the 2SC phase resemble that of the ordinary BCS theory, including vanishing resistivity and vanishing heat capacity because the quarks near the Fermi surface remain gapped.

\vskip 0.2cm
\item[(b)]    Phases with broken space symmetries, which are associated with a finite momentum of the condensate~\cite{Alford2008,Anglani2014} (hereafter FF phase) or deformation of the Fermi surface~\cite{Muther2003PhRvD} (hereafter DFS phase):
  \begin{eqnarray}\label{eq:QCD_FF} 
    &&\Delta_{2SC} \neq 0, \quad \delta\mu >\Delta/\sqrt{2},  \quad \vec P \neq 0  \quad \textrm{(FF)}\\
    &&\Delta_{2SC} \neq 0, \quad \delta\mu >\Delta/\sqrt{2},  \quad \delta\epsilon \neq 0  \quad \textrm{(DFS)}
  \end{eqnarray}
  where $\vec P$ is the center of mass momentum of a Cooper pair and $\delta\epsilon$ quantifies the quadrupole
  deformation of the Fermi surfaces of $u$ and $d$ quarks.

\vskip 0.2cm
\item[(c)]  Mixed phase(s)~\cite{Bedaque2003PhRvL}
  \begin{eqnarray}\label{eq:QCD_PS} 
\Delta_{2SC} \propto \langle \psi^T(x)C\gamma_5\tau_2\lambda_2\psi(x)\rangle \neq 0 , \quad  \delta\mu =0,\quad 0\le    x_s\le 1,  
  \end{eqnarray}
  which corresponds to a mixture between a perfectly symmetrical ``2SC'' superconductor and a normal system accommodating the excess number of  $d$ quarks. Here, $x_s$ is
  the filling factor defined as the ratio of the superconducting
  and total volumes. (We assume that there is an excess of $d$ over $u$ quarks, as is expected in
  quark matter in compact stars under $\beta$-equilibrium.)
 \end{itemize} 

 The color-flavor-locked (CFL) phase~\cite{Alford1999NuPhB} is expected to be the ground state of three-flavor quark matter at asymptotically
 large densities where the strange quark is massless, Fermi surfaces of quarks coincide, and, therefore, the pairing among quarks occurs in a particularly symmetrical manner.  At densities relevant for neutron stars, the perfect CFL phase is unlikely to be realized; rather,
  some of its variants have  $m_s\neq 0$ and/or $\delta\mu\neq 0$---chemical potential shifts between various flavors of
  quarks~\cite{Alford2005PhRvD}. Therefore, the phases listed above can be replicated with an allowance of additional non-zero $us$ and $ds$~pairings
  \begin{eqnarray}\label{eq:QCD_CFL} 
\Delta_{ud} \neq 0, \quad \Delta_{sd} \neq 0,  \quad \Delta_{su} \neq 0, \quad (m_s\neq 0;\, \delta\mu\neq 0). 
  \end{eqnarray}

A complete phase diagram of quark matter that includes most, if not all, of the phases mentioned above, is not available to date. However, various imbalanced superfluids, such as cold atoms, isospin asymmetrical nuclear matter, and flavor-imbalanced quark matter show a high degree of universality. Thus, possible structures of the phase diagram of quark matter can be conjectured by extrapolating
from the detailed studies of the phase diagrams of cold atomic gases~\cite{Strinati2018} and isospin asymmetrical nuclear matter~\cite{Stein2014PhRvC}. These are, clearly,  speculative and need to be confirmed using explicit computations of relevant quark phases.

\begin{figure}[H]
\includegraphics[width=0.49\hsize,angle=0]{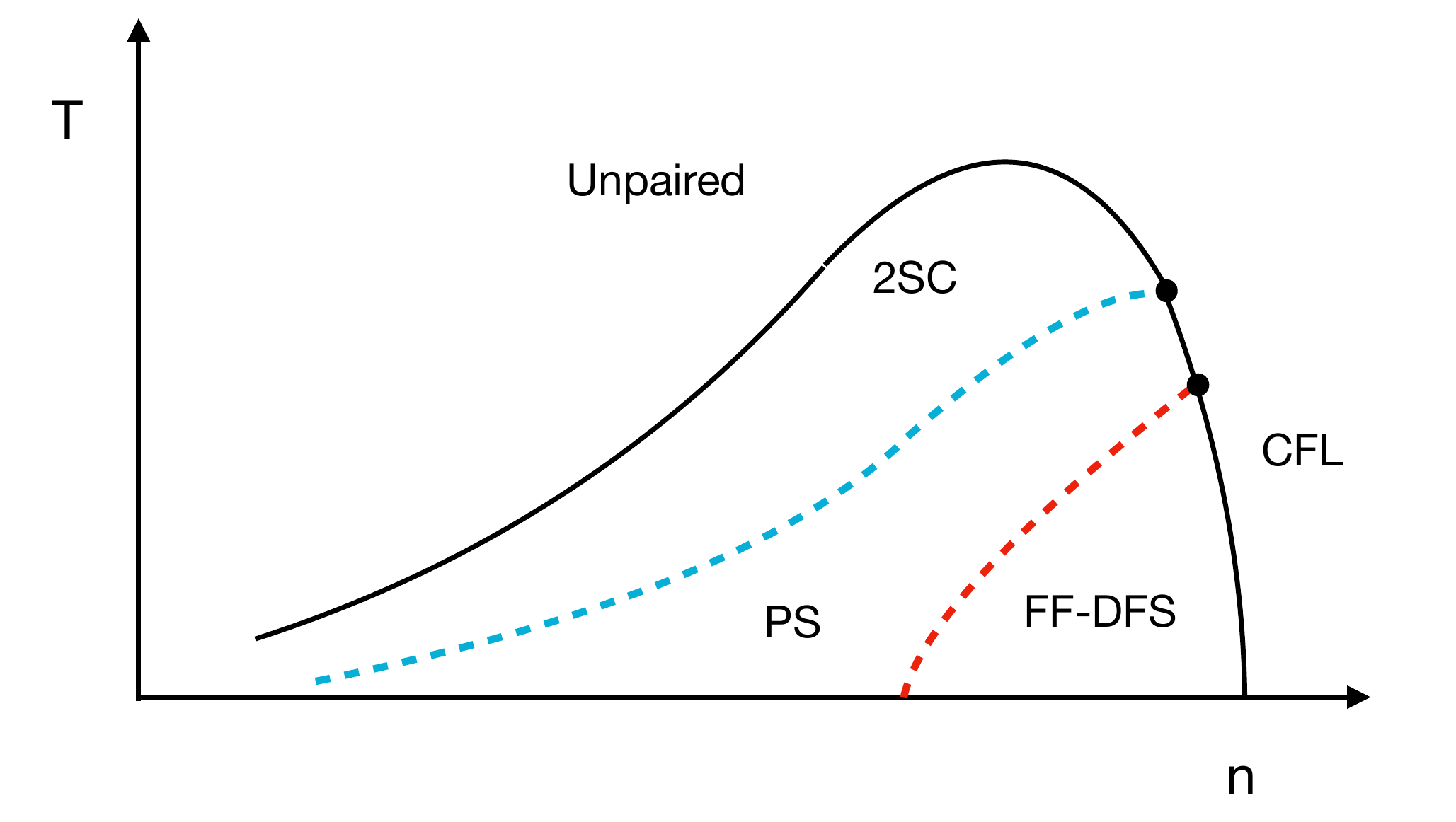}
\hskip 0.2cm
\includegraphics[width=0.49\hsize,angle=0]{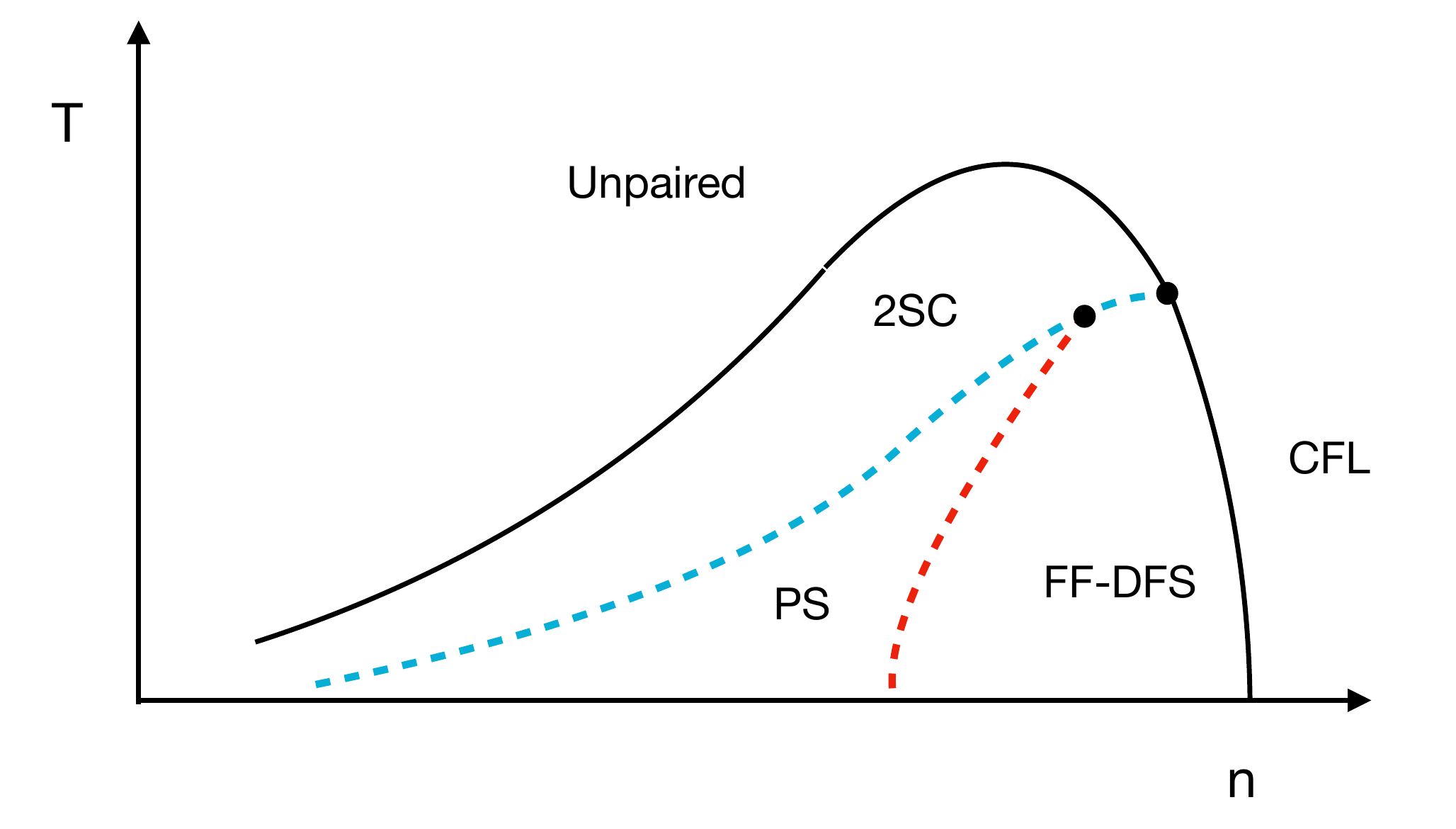}
\caption{Sketch 
 of the phase diagram of strongly interacting matter in the temperature
  and baryonic density plain, including (collectively indicated) modulated FF-phase and deformed Fermi surface DFS phase. The tri-critical points are shown
  with dots; the Lifshitz point is adjacent to the FF, unpaired/CFL phases, and homogenous (PS) phases.
  In the left panel, it is located on the unpairing or CFL-transition
  (solid line). The dashed lines correspond to the phase-separation lines among
various phases. Signatures of BCS--BEC  crossover/transition may emerge when moving from high to low densities.  }
\label{fig:2}
\end{figure}

Figure~\ref{fig:2} shows two schematic phase diagrams of color-superconducting matter in the density--temperature plane. For sufficiently large temperatures, the unpaired normal phase is the preferred state of matter, ignoring any other correlation beyond the pairing. The phases with broken symmetries, the FF and the DFS phases, are preferable in temperature--density strips at low temperatures and high densities. At lower temperatures, the PS phase is the preferred one. At higher temperatures, the spatially symmetric 2SC phase dominates. It is seen that the phase diagram contains two tri-critical points, i.e., the points where three different phases coexist.  The critical point, which has the FF state at the intersection, is a Lifshitz point as, per construction, it is a meeting point of the modulated (FF), ordered (PS/2SC), and disordered (unpaired) states. Of course, this is the case  if the transition temperature to the CFL phase is below the tri-critical temperature; otherwise, the unpaired state should be replaced by a variant of the CFL phase. Note that depending on the parameters of the model, two or one tricritical points may be located on the unpairing line or the line of transition to the CFL phase, as illustrated in Figure~\ref{fig:2}, left and right panels, respectively. The model can be tuned to produce a four-critical point if both points coincide. We also note that the low-density limit corresponds to the strong coupling regime where the pairs are tightly bound, whereas the high-density limit corresponds to the weak-coupling regime. Therefore, one can anticipate signatures of BCS--BEC crossover. These can be seen by examining several characteristic quantities, for example, the ratio of the coherence length to the interparticle distance $\xi/d$, where $\xi/d\gg1$ corresponds to the BCS  and $\xi/d\ll1$ corresponds to the BEC limit, or the ratio of the gap to the (average) chemical potential $\Delta/\mu$, where $\Delta/\mu \ll 1$ corresponds to the BCS and $\Delta/\mu \gg 1$ corresponds to the BEC limit. For discussions of BCS=-BEC crossover in dense quark matter, see Refs.~\cite{Sun2007,Sedrakian2009PhRvD,Ferrer2015NuPhA,Duarte2017PhRvD}. This phenomenon
shows a high degree of universality as well; see for example, the studies of nuclear matter~\cite{Lombardo2001PhRvC,Sedrakian2006PhRvC,Jin2010PhRvC}
and cold atoms~\cite{Giorgini2008RvMP,Strinati2018}.

 \section{Structure of Green Functions in Two-Flavor Quark Matter}
 \label{sec:GF}
 Order-by-order computations of the magnitude of the gaps in the
 superconducting phases can be carried out in the weak-coupling
 (extreme high-density) regime, where the one-gluon exchange is the
 dominant interaction.  Approximate Eliashberg-type equations for the
 flavor-symmetric 2SC phase were solved within one-gluon exchange
 approximation in Refs.~\cite{Son1999PhRvD,Schafer1999}, showing that
 the pairing gap scales with the coupling $g$ as
 \mbox{$\Delta\sim \mu g^{-5}\exp(-1/g)$.} Such a scaling also applies
 to the high-density CFL phase, where the perturbative approach is
 more reliable than at densities relevant to the 2SC phase.  More
 recently, Eliashberg-type equations were solved for
 two-flavor~\cite{Sedrakian2018PhLB} and three-flavor
 superconductors~\cite{AlfordWindisch2018}.  The first study used the
 quark--meson coupling model, keeping only the frequency dependence of
 the gap, whereas the second study kept frequency and momentum
 dependences but ignored the imaginary part of the pairing gap. These
 theories not only improve the description of quark matter but also
 lead to phenomenologically important implications, such as the
 presence of electrons in the CFL phase~\cite{AlfordWindisch2018},
 which are not allowed when the gap is
 constant~\cite{Rajagopal2001PhRvL}.

\begin{figure}[H]
\includegraphics[width=0.7\hsize,angle=0]{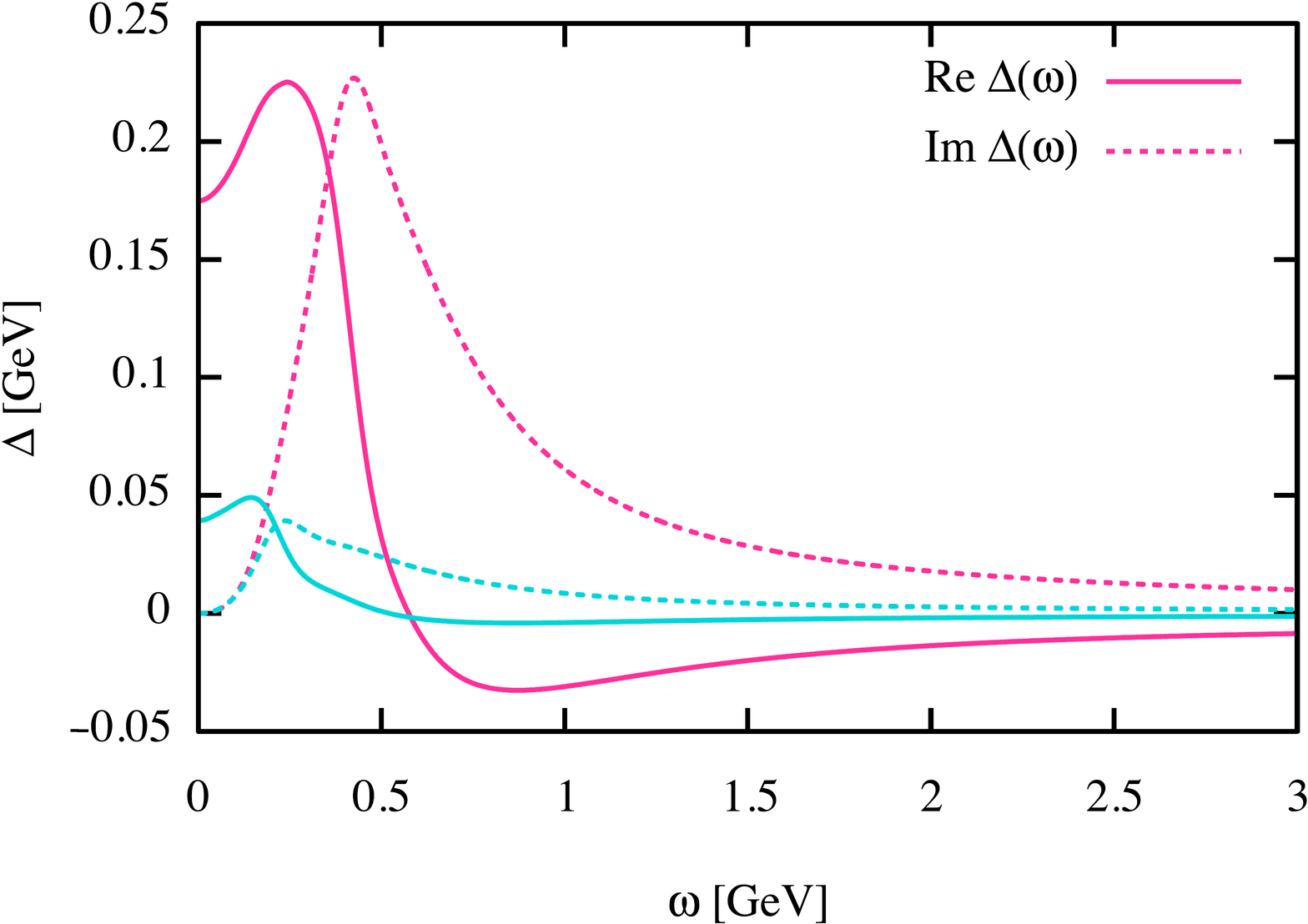}
\caption{Dependence of the real (solid) and imaginary (dashed)
  components of the positive energy projection of the gap function on
  frequency for two different values of the coupling shown by blue and
  red lines~\cite{Sedrakian2018PhLB}. The BCS theory predicts a
  constant on-shell value Re$\,\Delta =$ Const. and a vanishing
  Im$\,\Delta(\omega)$.}
\label{fig:3}
\end{figure}

We now briefly outline these approaches following Ref.~\cite{Sedrakian2018PhLB}. The inverse Nambu--Gorkov quark propagator is given by 
\be\label{eq:1.2}
S^{-1}(q)
= \left(\begin{array}{cc}{\slashed q}+{\mu}\gamma_0-m&\bar \Delta 
\\
                                 \Delta      & ({\slashed
q}-{\mu}\gamma_0+m)^T \end{array} \right),
\ee
where $q$ is the four-momentum and
$\Delta$ is the gap with $\bar\Delta \equiv \gamma_0\Delta^{\dag}\gamma_0$. Equation~\eqref{eq:1.2} is written for
the case of equal number densities of up and down quarks with a common chemical potential $\mu$ and mass $m$.
The bare quark--meson vertices $ \Gamma_\pi^i(q)$ and $\Gamma_\sigma(q)$ are given by
\bea
\label{eq:1.3b}
\Gamma_\pi^i(q)=  \left( \begin{array}{cc}\frac{\tau^i}{2}\gamma_5 &0 \\
                 0  &-(\frac{\tau^i}{2}\gamma_5)^T 
\end{array} \right),\quad
\label{eq:1.3c}
\Gamma_\sigma(q)= \left( \begin{array}{cc}\mathbb{I} &0 \\
                 0  &-\mathbb{I}
\end{array} \right),
\eea
where pions couple to quarks using a pseudo-scalar coupling, whereas
$\sigma$s couple via a scalar coupling, with $\mathbb{I}$ being a unit
matrix in the Dirac and isospin spaces.  Their propagators are given
by 
\bea\label{eq:1.6}
D_{\pi}(q)=\frac{1}{q_0^2-\vecq^2-m_\pi^2}, \qquad
D_{\sigma}(q)=\frac{1}{q_0^2-\vecq^2-m_\sigma^2},
\eea
where $m_{\pi/\sigma}$ values are the meson masses. 
The equation for the gap in  the Fock approximation  is  given via\vspace{-10pt}
\begin{adjustwidth}{-4.6cm}{0cm}
\bea\label{eq:1.7}
\Delta(k)
          &=& ig_\pi^2 \int\frac{d^4 q}{(2\pi)^4}
             \left(-\frac{\tau^i}{2}\gamma_5\right)^T S_{21}(q) 
             \frac{\tau^j}{2}\gamma_5\delta_{ij} D_\pi(q-k)\nonumber\\
             &+& ig_\sigma^2 \int\frac{d^4 q}{(2\pi)^4} (-\mathbb{I})^T 
S_{21}(q) \mathbb{I} D_\sigma(q-k) ,
\eea
\end{adjustwidth}
where $g_\pi$ and $g_\sigma$ are the coupling constants.  Adopting the color--flavor
structure of the gap function corresponding to a 2SC superconductor, one  then finds
\bea\label{eq:1.8}
\Delta^{ab}_{ij}(k)=(\lambda_2)^{ab}(\tau_2)_{ij}C\gamma_5
                    [\Delta_+(k)\Lambda^+(k)
                     +\Delta_-(k)\Lambda^-(k)], \quad 
\eea
where $a,b\dots$ refer to the color space, $i,j,\dots$ refer to the
flavor space, and the projectors onto the positive and negative states
are defined in the standard fashion as
 $\Lambda^{\pm}(k) = (E_k^{\pm} + \veca \cdot \veck + m\gamma_0) /2 E_k^{\pm}, $ where $E_k^{\pm} = \pm\sqrt{\veck^2+m^2}$ and $\veca =\gamma_0\gammavec$. The coupled Equations~\eqref{eq:1.2}--\eqref{eq:1.8} must be solved for the gap function, which is a function of three-momentum and the frequency. In the low-temperature limit, the relevant momenta are close to the Fermi momentum and the dependence on the magnitude of the three-momentum can be eliminated by fixing it at the Fermi momentum. The gap Equation \eqref{eq:1.7} then depends only on the energy, which reflects the fact that the pairing interaction is not instantaneous---a common feature of the Fock self-energies in ordinary many-body perturbation theory. The solutions for the positive energy projection of the gap function are shown in Figure~\ref{fig:3} as a function of frequency. The structure of the real and imaginary components of the gap function shows a maximum around frequencies at which the meson spectral functions are peaked. Thus, it is important to include the retardation effect when the color superconductor is probed at such frequencies. In the low-frequency limit, it is sufficient to use the BCS approximation where the interaction is instantaneous so that the imaginary part vanishes Im$\,\Delta(\omega)=0$ and the real part is a constant Re$\,\Delta =\Delta(\omega=0)$.

Ref.~\cite{AlfordWindisch2018} considered the full momentum and energy dependence of the gap in the Fock approximation within the Yukawa model but  neglected the imaginary part of the anomalous self-energy. Their work shows that the retardation implies a CFL phase that is not a perfect insulator, as charge neutrality requires some electrons to be present in matter. This is not the case in the treatment based on the BCS model~\cite{Rajagopal2001PhRvL}. Thus, the phenomenology of the CFL phase is modified: its specific heat, thermal conductivity, and magnetic response will change due to the contribution of electrons.  This example, in which the simple BCS ansatz for the gap is replaced by a more complete gap function, demonstrates some unexpected features of color superconductors, which may be important for their transport and dynamic response to various probes.

\section{Equation of State and Mass--Radius diagram}
\label{sec:MR}
We have seen in the previous section that the phase diagram of quark matter may have a complicated structure. At minimum, there are two robust phases of color superconducting matter: the low-density two-flavor 2SC phase and the high-density three-flavor CFL phase. See, for example, Ref.~\cite{Bonanno2012} for a Nambu--Jona-Lasinio study and compact stars with two phase transitions in this model.  However,  additional phases are very likely because it is energetically
favorable to break the rotational and translational symmetries due to the stress on the paired state induced by the finite mass of the strange quark and $\beta$-equilibrium, which induces
disparity in the chemical potentials of $u$ and $d$ quarks. In addition, or alternatively, quarkyonic phases may interfere.

For the specific computation below, we adopt a covariant density functional EoS of nuclear matter in the nucleonic phase~\cite{Lijj2019b,Lijj2020a,Lijj2021}.  This EoS, in the absence of the
phase transition to quark matter, produces nucleonic compact stars with a maximum $m\equiv M/ M_{\odot}\simeq 2.6$, where $M$ is the gravitational mass of the star and $M_{\odot}$ is the solar mass.
Allowing for phase transition to quark matter, we consider a straightforward extension of the constant speed of sound EoS of Ref.~\cite{Alford2017PRL} that allows for  a conformal phase of quark matter  at high densities with a constant speed of sound, i.e.,  
\bea\label{eq:EoS}
p(\ep) =\left\{
\begin{array}{ll}
p_{1}, &  \ep_1 < \ep < \ep_1\!+\!\Delta\ep_1 \\[0.5ex]
p_1 + s_1 \bigl[\ep-(\ep_1\!+\!\Delta\ep_1)\bigr],
          & \ep_1\!+\!\Delta\ep_1 < \ep < \ep_2 \\[0.5ex]
p_2,   & \ep_2 <\ep < \ep_2\!+\!\Delta\ep_2   \\[0.5ex]
             p_2+ s_2\bigl[\ep-(\ep_2\!+\!\Delta\ep_2)\bigr],
         & \ep_2\!+\!\Delta\ep_2 < \ep < \ep_3\\
p_3, & \ep_3 <\ep < \ep_3\!+\!\Delta\ep_3 \\[0.5ex]
p_3+ s_{3}\bigl[\ep-(\ep_3\!+\!\Delta\ep_3)\bigr], &  \ep > \ep_3
\end{array}
\right.
\eea
where the three pairs of  the pressure and energy density $p_{1,2,3}$ and $\ep_{1,2,3}$
correspond, respectively, to the transition from hadronic to quark matter, from a 
low-density (2SC) quark phase  to a high-density (CFL) quark phase, and
from the high-density quark phase to the conformal fluid.
The squared sound speeds in the quark phases are denoted by  $s_1$, $s_2$, and $s_3
=c^2_{\rm conf.}$. Note that we assume that the 2SC and CFL quark phases are separated by a jump at the phase boundary, as it follows from the study of Ref.~\cite{Bonanno2012}. At high densities, the CFL 
phase reaches the ``conformal limit''  where the interactions are dominated by the underlying conformal symmetry of QCD. In this limit, the speed of sound squared is $s_3= 1/3$ (in units of speed of light), whereas the effects of the pairing gap of the CFL phase can be neglected in a first approximation. 
Note that we allow for a small jump between proper CFL and conformal zero-gap fluid, but its effect on the observables is marginal, i.e., a smooth interpolation would not change the results. 


According to Equation~\eqref{eq:EoS}, the modeling of the EoS of quark phases
involves the following parameters:

\begin{itemize}
\item The three (energy) densities at which the sequential transitions between the
  nucleonic phase, 2SC phase,  CFL phase, and conformal fluid take place.

\item The magnitudes of the jumps in the energy density at the points of 
  the transition from nuclear to the 2SC phase,
  $\Delta\varepsilon_{1}$, from the 2SC 
  to the CFL phase, $\Delta\varepsilon_{2}$, and from the CFL to the conformal fluid phase
  $\Delta\varepsilon_{3}$.

\item The speeds of sound in the 2SC and CFL phases
  $s_{1}$ and $s_{2}$. The speed of sound of the conformal fluid is held fixed at 
  $s_3 = 1/3$. Note that for any phase, $s\le 1$ by causality.
\end{itemize}

Our model EoS is constructed using the following parameters. The transition pressure
and energy density from nuclear and quark matter are $p_1=1.7\times 10^{35}$ dyn/cm$^2$
and $\ep_1 =8.4\times 10^{14}$ g cm$^{-3}$, respectively. The magnitude of the first jump 
$\Delta\ep_1= 0.6~\ep_1$. The upper range of the energy density of the 2SC phase is
determined as $\ep_1^{\rm max}=\delta_{2SC}(\ep_1+\Delta\ep_1)$, where $\delta_{2SC}$ is
a dimensionless parameter measuring the width of the 2SC phase. The magnitude of
the second jump is parametrized in terms of the ratio parameter \mbox{$r=\Delta\ep_2/\Delta\ep_1$}. The extent of the CFL phase is determined by limiting its energy density range to \mbox{$\ep_2^{\rm max}=\delta_{CFL}(\ep_2+\Delta\ep_2)$}. The transition to the conformal fluid is assumed to be of the first order with a small (compared to other scales) energy-density
jump equal to $0.1r$. The transition to the conformal fluid phase occurs at densities
$\ep_3 = 2.25$--$2.57\times 10^{15}$ g cm$^{-3}$, i.e.,  by about a factor of 10 larger than the saturation density. The speeds of sound squared are fixed as
\bea
s_1 =0.7,\qquad s_2=1, \qquad s_3=\frac{1}{3}.
\eea

The values of $s_1$ and $s_2$ are chosen to obtain triplet configurations with large enough masses of hybrid stars. The magnitudes of jumps between the nuclear, 2SC, and CFL phases were chosen suitably to produce twin and triplet configuations~\cite{Alford2017PRL}.

\begin{figure}[H]
\includegraphics[width=0.8\hsize,angle=0]{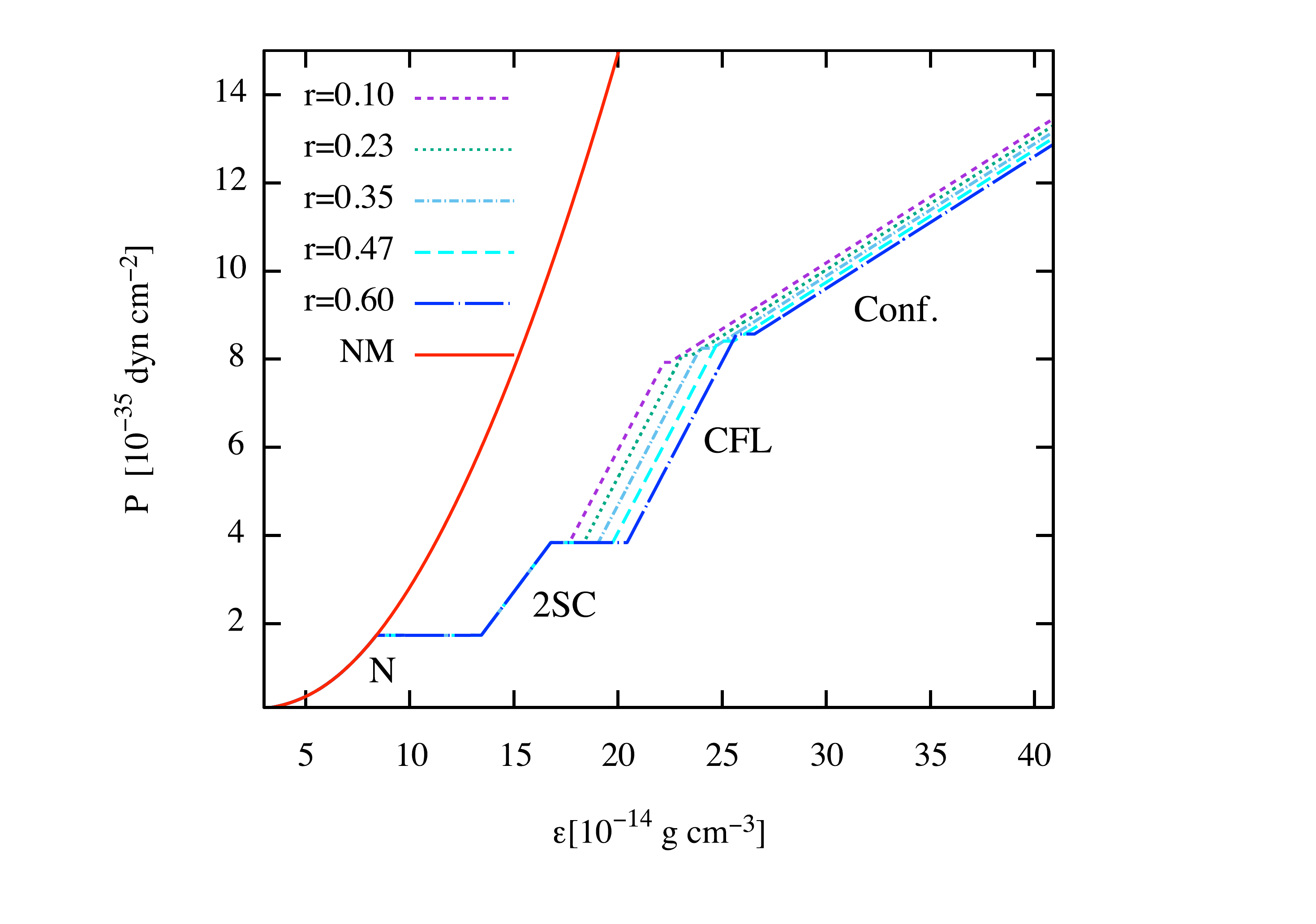}
\caption{The pressure vs. energy density (EoS) for nucleonic matter
  (long-dash-dotted curve) and a series of EoSs that contain two
  sequential phase transitions via Maxwell construction manifest in
  the jumps of the energy density. The models differ by the magnitude
  of the second jump measured in terms of the ratio
  $r=\Delta\epsilon_2/\Delta\epsilon_1$.}
\label{fig:EoS}
\end{figure}

\begin{figure}[H]
\includegraphics[width=0.85\hsize,angle=0]{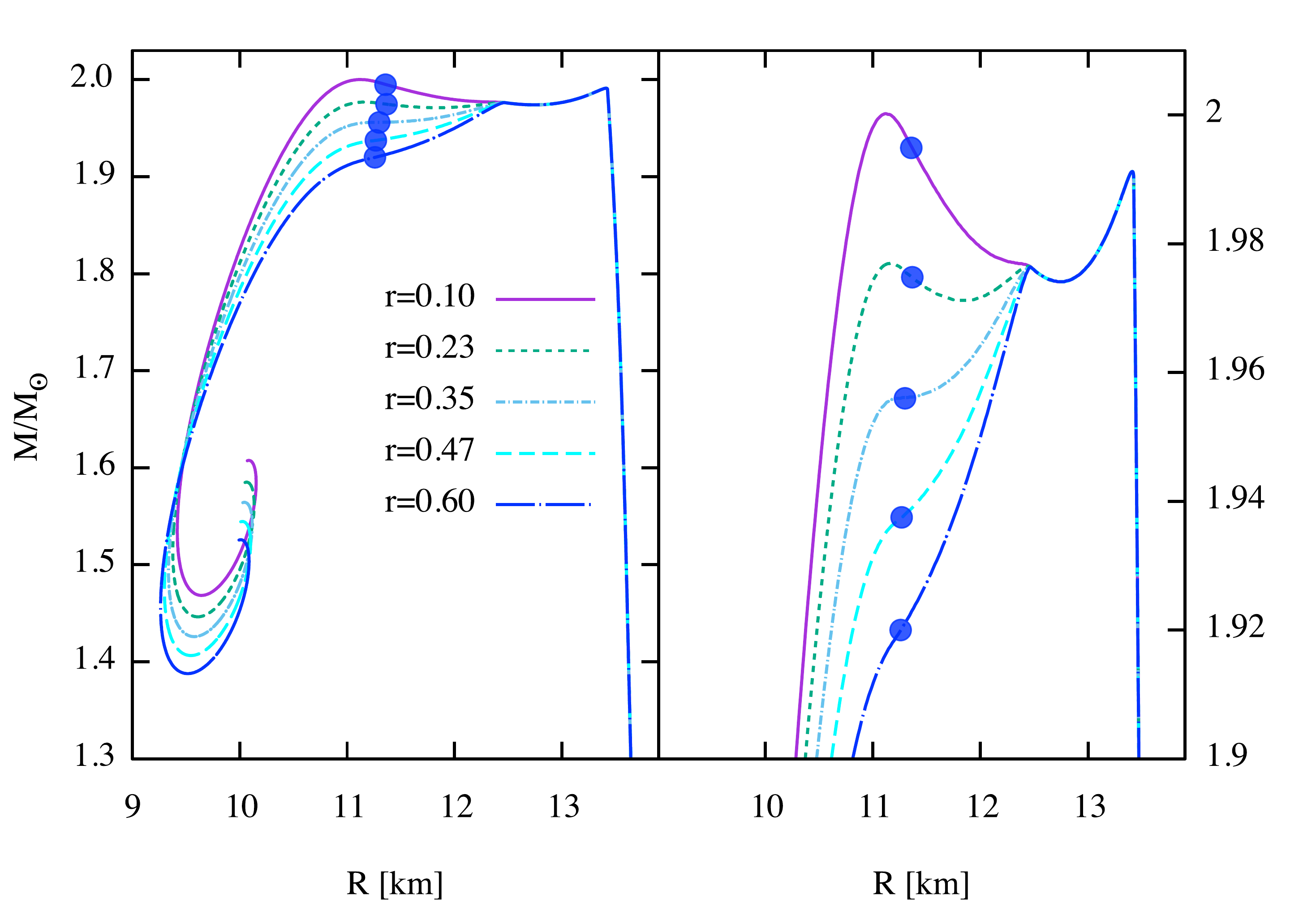}
\caption{The $M$--$R$ relations corresponding to the EoS shown in
  Figure~\ref{fig:EoS} for several ratios of the second jump. The
  right panel enhances the high-mass range to demonstrate the
  emergence of the triplets and the fourth family of compact
  stars. Note that the different MR curves cross each other at the
  special point located in the low-mass and low-radius region, in
  analogy to the single-phase transition case; see
  Ref.~\cite{Cierniak2020EPJST}. The blue circles indicate the stars
  in which the central density corresponds to $10~n_{\rm sat}$ at
  which the conformal fluid sets in.}
\label{fig:MR}
\end{figure}

Figure~\ref{fig:EoS} shows a collection of EoS constructed based on Equation~\eqref{eq:EoS}, which 
shares the same low-density nuclear EoS. In this collection, we vary the parameter
$r$ (as indicated in the plot) for fixed values $\delta_{2SC}=\delta_{CFL}=0.27$. The corresponding  $M$--$R$ relations for static, spherically symmetrical stars obtained from solutions to Tolman--Oppenheimer--Volkoff equations are shown in Figure~\ref{fig:MR}.

For the chosen magnitude of the first jump $\Delta\ep_1$, the $M$--$R$ curves show the phenomenon of twins---two stars of the same mass but different radii. The radii of twins differ by about 1~km. The more compact configuration is a hybrid star, i.e., a star with a quark core and nuclear envelope, whereas the less compact counterpart is a purely nucleonic star. The second phase transition may or may not result in a classically stable sequence depending on the value of the parameter $r$ parameterizing the magnitude of the second jump. For small jumps $r=0.1$ and 0.23, new stable branches arise, which are continuously connecting to the stable 2SC branch ($r=0.1$) or are separated by a region where the stars are unstable ($r=0.23$). It is seen that, in this case, triplets of stars with different radii but the same masses appear. The densest stars contain, in addition to the 2SC phase,  a layer of the CFL phase, whereby the central density on the stable branch can exceed the onset density of the conformal fluid.  This implies that the densest member of a triplet will contain in its center conformal fluid with $c_{\rm conf.}=1/\sqrt{3}$. For each $M$-$R$ curve in Figure~\ref{fig:MR}, the star with a  central density at which the conformal fluid first appears is shown by a dot (this density is fixed at $10~n_{\rm sat}$). The stable branch of conformal fluid containing stars is followed by a
classically unusable branch with $dM/d\rho_c <0$. For asymptotically large central densities, the masses and radii increase again. The family of the EoSs that  differ only in the value of the parameter $r$ cross at a ``special point''. This type of crossing has been observed for twin star configurations with a variation in a particular parameter of the EoS~\cite{Cierniak2020EPJST}; however, the EoS excluded two sequential phase transitions. The behavior of $M$--$R$ curves at very high central densities differs from the ones that were found in Ref.~\cite{Li2023PhRvD}, where a branch of ultracompact twin stars with masses of the order of $1~M_{\odot}$ and radii in the range of $6$--$7$~km were found for a single phase transition from the nuclear matter to the quark phase.  Thus, we conclude that the
high-density asymptotics of the EoS modifies the behavior of the $M$--$R$ curves if the conformal
limit is achieved at densities of the order of $10~n_{\rm sat}$.
\vspace{-6pt}

The observation above may have phenomenological implications for the following reason.  The stability of stellar configurations is commonly determined by the requirement that the star's mass must increase with increasing central density (or central pressure), i.e., $\partial M/\partial\rho_c>0$.  An alternative and physically more transparent method is to compute the radial modes of oscillation of a star and determine the stable configurations from the requirement that their frequencies are real.  Ref.~\cite{Pereira2018} showed that the classical stability conditions fail if the conversion rate is slow, i.e., if its characteristic timescale is longer than the period of oscillations. In that case, the fundamental modes are stable even when $\partial M/\partial\rho_c<0$; i.e., stars with central densities larger than the one corresponding to the maximum-mass star (which lie to the left from the maximum on the $M$--$R$ diagram in Figure~\ref{fig:MR}) will be stable. This observation also applies to configurations with two-phase transitions, as shown in Refs.~\cite{Goncalves2022,Rau2022}. Furthermore, Ref.~\cite{Rau2022} shows that, in this case, the classically unstable stars contribute to the count of same-mass stars, which leads to the appearance of higher-order multiplets such as quadruplets, quintuplets, and sextuplets.  We will return to the stability of hybrid stars in Section~\ref{sec:modes}.

\section{Cooling of Compact Stars with Quark Matter Cores}
\label{sec:cooling}

The cooling of compact stars may provide indirect information about quark phases in hybrid stars.
The properties of phases of dense quark matter affect both neutrino emission and the specific heat content that determine the cooling rate of a compact object in general; see
Refs.~\cite{Kyra2023PhRvC,Zapat2022,Sedrakian2017EPJWC,Sedrakian2016EPJA,deCarvalho2015PhRvC,Grigorian2015,Noda2013ApJ,Negreiros2012PhRvC,Hess2011PhRvD,Sedrakian2013AandA,Stejner2009ApJ,Blaschke2007PrPNP,Anglani2006PhRvD,Popov2006PhRvC,AlfordJotwani2005PhRvD,Shovkovy2003,Blaschke2001}.

Non-superconducting relativistic quark matter cools predominantly via the direct Urca processes involving $d$, $u$, and $s$ quarks~\cite{Duncan1984ApJ}
\bea
\begin{aligned}
& d \rightarrow u+e+\bar{\nu}_e ,\\
& u+e \rightarrow d+\nu_e, \\
& s \rightarrow u+e+\bar{\nu}_e, \\
& u+e \rightarrow s+\nu_e,
\end{aligned}
\eea
where $\nu_e$ and $\bar\nu_e$ are the electon neutrino and antineutrions.
The neutrino emissivity through the direct Urca process for non-strange quarks is given via~\cite{Iwamoto1980PhRvL}
\bea\label{eq:epsilon1}
\epsilon_{\beta}=8.8 \times 10^{26} \alpha_s\left(\frac{n}{n_{\text {sat }}}\right) Y_e^{1 / 3} T_9^6 \quad \mathrm{ergs} ~\mathrm{cm}^{-3} ~\mathrm{~s}^{-1},
\eea
where $n$ is the baryon density, $Y_e$ is the electron fraction, $T_9$ is the temperature in units of $10^9 \mathrm{~K}$, and $\alpha_s$ is the running strong coupling constant. The emissivity given by Equation~\eqref{eq:epsilon1} implies that the stars containing unpaired quark matter would cool quickly via this direct Urca process.  The cooling would be slower if the quark spectrum contains a gap. In
the case of the phenomenologically relevant 2SC phase, two alternatives are possible, depending on whether the Fermi surfaces of quarks are a full gap or they contain zero-gap segments (nodes).  The latter feature arises in the case of pairing between fermions on different Fermi surfaces, as discussed in Section~\ref{sec:phases}.

Ref.~\cite{Jaikumar2006} studied a generic case where the quark spectrum is gapped if the parameter $\zeta=\Delta_0 / \delta \mu$ associated with the new scale $\delta \mu=\left(\mu_d-\mu_u\right) / 2$, where $\mu_{u, d}$ are the chemical potentials of light quarks and $\Delta_0$ is the gap for $\delta \mu=0$.  The suppression of emissivity by pairing is qualitatively different in the cases in which $\zeta>1$ and $\zeta<1$.  The novelty arises in the second case, where Fermi surfaces have nodes and particles can be excited around these nodes without any energy cost (which is not the case for gapped Fermi surfaces). Note that in the case of the FF phase, the shift in the chemical potential is replaced by a more general function---the anti-symmetric in the flavor part of the single particle spectrum of up and down quarks. This new physics can be captured by adopting a generic parameterization of the suppression factor of the quark Urca process with pairing suggested in Ref.~\cite{Jaikumar2006}. The neutrino emissivity
of the 2SC phase $\epsilon_{2SC}^{r g}$ can be related to the Urca rate in the normal phase \eqref{eq:epsilon1} as
\bea
\epsilon_{2SC}^{r g}\left(\zeta ; T \leq T_c\right)=2 f(\zeta) \epsilon_\beta, \quad f(\zeta)=\frac{1}{\exp \left[(\zeta-1) \frac{\delta \mu}{T}-1\right]+1} ,
\eea
where the parameters $\zeta$ and $\delta\mu$ were introduced above,
$T$ is the temperature, and $T_c$ is the critical temperature of the
phase transition from normal to the 2SC phase.  Furthermore, the
parameter $\zeta(T)$ is temperature-dependent and we adopt the
parametrization
\bea
\zeta(T)=\zeta_i-\Delta \zeta g(T),
\eea
where $\zeta_i$ is the initial value, $\Delta \zeta$ is the constant change in this function, and the function $g(T)$ describes the transition from the initial value $\zeta_i$ to the asymptotic final value $\zeta_f=\zeta_i-\Delta \zeta$. The transition is conveniently modeled by the following function\vspace{6pt}
\bea\label{eq:gT}
g(T)=\frac{1}{\exp \left(\frac{T-T^*}{w}\right)+1},
\eea
which allows one to control the temperature of transition by adjusting the parameter $T^*$ and the smoothness of the transition via the width parameter $w$. An additional issue to address is the role of the blue quarks that  do not participate in the 2SC pairing. Blue quarks may pair among themselves due to the attractive component of the strong force as in the ordinary BCS case (as both members of the Cooper pair are on the same Fermi surface). Then, the emissivity of blue quarks in the superfluid state is given by 
\bea
\epsilon_{BCS}^b\left(T \leq T_{cb}\right) \simeq \epsilon_{\beta}^b\left(T>T_{cb}\right)\exp \left(-\frac{ \Delta_b}{T}\right),
\eea
where $\Delta_b$ is the gap in the blue quark spectrum, $T_{cb}$ is the corresponding critical temperature, and $\epsilon_{\beta}^b$ is the neutrino emissivity of blue quarks in the normal state. 
As discussed in Section~\ref{sec:MR}, the densest members of the triplets contain cores of CFL matter
that is fully gapped.
In this case, the excitations are the Goldstone modes of the CFL phase. Their 
emissivity, as well as the specific heat, is rather small compared to other phases due to their
very small number density~\cite{Jaikumar2002PhRvD}.  In the following discussion, we will ignore the role of the CFL phase in the cooling of hybrid stars. In the conformal fluid phase, we expect three-flavor
pairing gap $\Delta\sim \mu g^{-5}\exp(-1/g)$, $g=\sqrt{4\pi\alpha_s}$, with a spin--flavor structure of the CFL phase.  

\begin{figure}[H]
\includegraphics[width=0.85\hsize,angle=0]{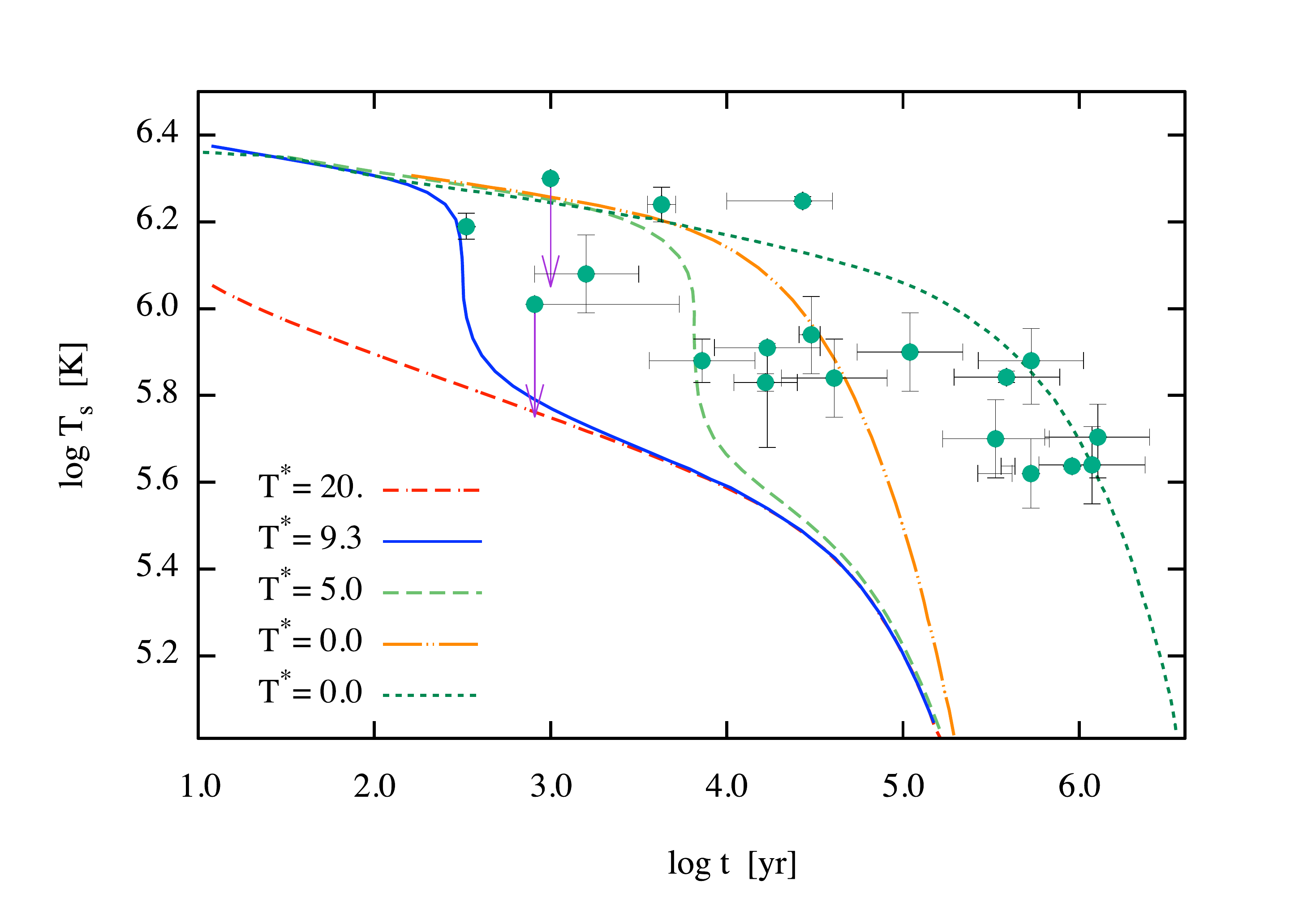}
\caption{Cooling tracks of compact stars with quark cores in the
  surface-temperature--age diagram.  The masses of the stars are the
  same, $M = 1.93~M_{\odot}$, and the different curves correspond to
  different values of the parameter $T^*$ in units of keV, except the
  dotted line, which corresponds to 1~$M_{\odot}$ mass nucleonic
  compact star without a quark core. The observational points with
  error bars are shown by green circles; the arrows show the upper
  limits on surface temperatures of known objects. }
\label{fig:Cooling}
\end{figure}

Let us turn to the cooling simulations of hybrid stars with a gapless 2SC superconductor.  The cooling tracks are shown in Figure~\ref{fig:Cooling}, and the input physics beyond the emissivities is discussed elsewhere~\cite{Hess2011PhRvD,Sedrakian2013AandA,Sedrakian2016EPJA,Sedrakian2017EPJWC}.  The key parameter regulating the behavior of the cooling curves in Figure~\ref{fig:Cooling} is the temperature $T^*$, which controls the transition from the gapped to ungapped 2SC phase. Similar results were obtained in the context of rapid cooling of the compact star in Cassiopeia (Cas) A remnant in Ref.~\cite{Sedrakian2013AandA,Sedrakian2016EPJA,Sedrakian2017EPJWC}. The model has a second parameter, the gap for blue-colored quarks $\Delta_b$, which prohibits rapid cooling via the Urca process involving only blue quarks. The  third parameter $w$ in Equation~\eqref{eq:gT} 
accounts for the finite time scale of the phase transition---see Refs.~\cite{Sedrakian2013AandA,Sedrakian2016EPJA}---but it is important only for the fine-tuning of the cooling curves close to the age of the Cas A.
The various cooling tracks shown in Figure~\ref{fig:Cooling}  correspond to various
values of $T^*$  for fixed values of $w$ and
$\Delta_b$ and stellar configuration of mass $1.93~M_{\odot}$.
It is seen that if $T^*$ is small, then the quark core does not influence the cooling, because during the
entire evolution $T>T^* $; therefore the neutrino emission is suppressed by the fully gapped
Fermi surfaces of red-green quarks. For large   $T^*$, early transition to the gapless phase occurs, and the star cools fast via the direct Urca process. Note that the value of $T^*$ can be fine-tuned to reproduce not only the current temperature of Cas A but also the fast decline claimed to be observed during the last decade or so; see Ref.~\cite{Shternin2023MNRAS} and references therein.
From the brief discussion above, one may conclude that the phase transitions within
the cold QCD phase diagram may induce interesting and phenomenologically relevant
changes in the cooling behavior of compact stars.
Although we will not discuss in any depth the dependence of cooling tracks on the stellar mass,
it should be pointed out that the onset of new phases in the interiors of compact stars, for example, hyperonization, meson condensation, and phase transition to quark matter, lead to mass hierarchy in the cooling curves~\cite{Raduta2018MNRAS,Raduta2019MNRAS,Anzuini2022MNRAS,Tsuruta2023ApJ}.
Typically, one finds that heavier stars that have central densities beyond the threshold for the onset of the new phase cool faster than the light stars containing only nucleonic degrees of freedom. This is also the case for models of stars studied here. For example, stars with masses $M\sim 1.1--1.6\, M_{\odot}$ remain warm over longer time scales and are thus hotter than their heavy analogs, which develop large quark cores.

\section{Stability Criteria for Hybrid Stars}
\label{sec:modes}

The oscillation modes of a compact star are important probes of their internal structure, as has been shown in the case of $g$ modes, which are sensitive to the size of the density jumps at a first-order phase transition between hadronic and quark matter~\cite{Orsaria2019,Wei2020,Jaikumar2021}.  They are expected to leave an imprint on the emitted gravitational wave signal during the binary inspiral of a neutron star, as well as in the post-merger phase~\cite{Bauswein2019,Weih2020,Liebling2021,Prakash2021}.

As discussed briefly in Section~\ref{sec:MR}, high-central density stars on the descending branch of $M$--$R$ diagram can have phenomenological implications if they are stabilized by some mechanism, which we discuss in this section.  The main mode of instability for non-rotating, spherically symmetrical fluid stars in general relativity is the instability against the radial $f$-mode of oscillations~\cite{Chandrasekhar1964ApJ}.  If the $f$-mode frequency $\omega_f^2>0$, the stellar configuration is dynamically stable, and it is unstable if $\omega_f^2<0$. The location of this instability point on the $M$-$R$ diagram agrees well with the turning point of the mass--central-density ($M-\rho_c$) curve. The stars on the ascending branch are stable, whereas those on the descending branch are unusable. The maximum mass is the point of marginal stability.  Numerical simulations found some violations of this criterion~\cite{Gourgoulhon1995,Takami2011MNRAS}, but quantitative deviations are insignificant.  However, recent work found that the agreement between these criteria is strongly violated for stars with first-order phase transition, as we review below.

Early work on stellar oscillations with phase transitions inside the star was carried out in the Newtonian theory assuming uniform phases~\cite{Bisnovatyi-Kogan1984,Haensel1989}. Two possibilities arise depending on the interplay between the scales in the problem: (a) when the conversion rate from one phase to another is fast, the interface between phases oscillates as a whole when perturbed; (b) if, however, conversion is slow, then the interface is fixed over the period of characteristic oscillations. The second case is interesting because, as shown in Ref.~\cite{Pereira2018}, the sign of $\omega_f^2$ does not change at the maximum mass $M_{\rm max}$ but stays positive over a segment where $\partial M/\partial\rho_c<0$. This implies that the classically unstable branch becomes stable against $f$-mode oscillations. Several subsequent studies confirmed this feature in the case of single-~\cite{Curin2021} and two-phase transitions~\cite{Goncalves2022,Rau2022}. The case of two-phase transition was extended in several directions in Ref.~\cite{Rau2022} by focusing on EoS, which supported classical twin and triplet star configurations, as discussed in Section~\ref{sec:phases}. It was shown that in the case of slow conversion, higher-order multiplet stars arise, since now the stars on the $\partial M/\partial\rho_c<0$ segments of the mass--central-density curve are located on the stable branch. Also, the properties of the reaction mode of a compact star~\cite{Haensel1989}, which arises in  case (a) with one or more rapid phase transitions, were studied.

The fundamental modes of hybrid stars are obtained from the set of equations~\cite{Chanmugam1977,Gondek1997}
\begin{eqnarray}
\frac{\textrm{d}\xi}{\textrm{d}r}&=&\left(\frac{\textrm{d}\nu}{\textrm{d}r}-\frac{3}{r}\right)\xi-\frac{\Delta P}{r\Gamma P},
\label{eq:XiEq}\\
\frac{\textrm{d}\Delta P}{\textrm{d}r}&=&\left[\textrm{e}^{2\lambda}\left(\omega^2 \textrm{e}^{-2\nu}-8\pi P\right)+\frac{\textrm{d}\nu}{\textrm{d}r}\left(\frac{4}{r}+\frac{\textrm{d}\nu}{\textrm{d}r}\right)\right]
\left(\rho+P\right)r\xi\nonumber\\
&-&\left[\frac{\textrm{d}\nu}{\textrm{d}r}+4\pi(\rho+P)r\textrm{e}^{2\lambda}\right]\Delta P,
\label{eq:DeltaPEq}
\end{eqnarray}
where $\xi=\xi_{\rm dim}/r$, with  $\xi_{\rm dim}$ being  the Lagrangian displacement, $r$ the radial coordinate, $\Delta P$ the Lagrangian perturbation of pressure,  $\rho$ the mass--energy-density, $\omega$  the angular frequency, $\Gamma$  the adiabatic index, and $\text{e}^{2\nu}$ and $\text{e}^{2\lambda}$ the metric coefficients entering the Tomann--Oppenheimer--Volkoff equations.
In a first approximation, the  adiabatic index for a chemically equilibrated relativistic fluid can be taken as that of the matter in $\beta$-equilibrium
$
\Gamma = [(\rho+P)/P]\left(dP/d\rho\right).
$
The set of \mbox{Equations \eqref{eq:XiEq} and \eqref{eq:DeltaPEq}} can be solved provided the boundary conditions are known. These are specified by assuming that the displacement field is divergence-free at the center and that the Lagrangian variation of the pressure vanishes at the surface of the star:
\begin{equation}
\Delta P(r=0) = -3\Gamma P\xi(r=0), \qquad \Delta P(r=R)=0.
\end{equation}

The $\omega^2$ values obtained in this manner are usually labeled according to the number of radial nodes in $\xi$ and the $f$ mode corresponds to the nodeless mode.

In the case of multiple phase transitions in the QCD phase diagram,
one needs junction conditions that relate the values of Lagrangian
perturbations on both sides of the interface between phases. Such
junction conditions already appear in the work of
Ref.~\cite{Haensel1989} in the Newtonian cases, whereas the the
general relativistic case is treated in Ref.~\cite{Karlovini2004}.
For the {\it slow conversion rate } one has the junction condition
\begin{equation}
\left[\Delta P\right]^+_-=0, \qquad \left[\xi\right]^+_-=0;
\label{eq:SlowConversionJunctions}
\end{equation}
for {\it rapid conversion rate}, one has
\begin{equation}
\left[\Delta P\right]^+_-=0, \qquad \left[\xi-\frac{\Delta P}{r}\left(\frac{\textrm{d}P}{\textrm{d}r}\right)^{-1}\right]^+_-=0,
\label{eq:RapidConversionJunctions}
\end{equation}
where $+/-$ refer to the high- and low-density sides of the transition, respectively.
At present, it is not possible to state with confidence which limit is realized in quark
matter, as the conversion rate varies significantly over the parameter space; see Ref.~\cite{Bombaci2016} for a discussion and earlier references. Ref.~\cite{Rau2022} considered
modified junction conditions that smoothly interpolate between the two limiting cases.

\begin{figure}[H]
\includegraphics[width=0.7\hsize,angle=0]{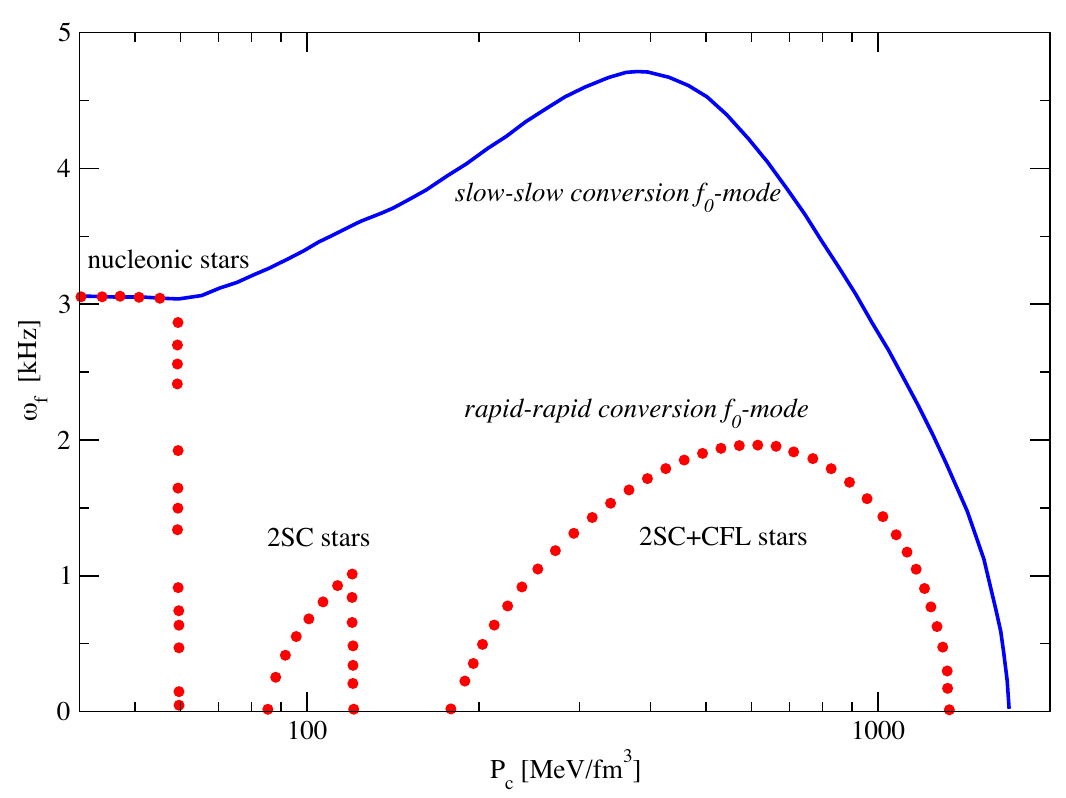}
\caption{The fundamental mode of triplet stars as a function of the
  central pressure of the configuration in the cases when both
  nucleonic-2SC and 2SC-CFL interfaces feature slow (solid line) or
  rapid conversion (dotted line). In the case of rapid conversion, the
  classical stability criteria apply; i.e., there are no real $f_0$
  solutions in the region where stars are unstable.  The corresponding
  curves are not shown.  For more details, see Ref.~\cite{Rau2022}.}
\label{fig:Modes}
\end{figure}

Phenomenologically, the most interesting implication of the modified stability criteria is the existence of new stable configurations beyond those that are classically stable. In particular, in the case where twins and triplets exist according to classical criteria of stability, additional configurations will arise when conversion between phases at the interface is slow.  These can form quadruplets (the maximum number in the case of twins) and quintuplets and sextuplets in the case of triplets. A particular case that allows for classical triplet stars is illustrated in Figure~\ref{fig:Modes}, adapted from Ref.~\cite{Rau2022}.  The fundamental mode frequency $\omega_f$ is shown as a function of the central pressure of the stars in two cases when both interfaces (i.e., nucleonic to 2SC and 2SC-CFL) feature rapid or slow conversion. (The case of rapid--slow and slow--rapid conversions are intermediate cases, and we omitted them.) To recover the classical case, one needs to assume that the conversion at each interface is rapid: in this case, the instability region is characterized by the vanishing of the real part of $\omega_f$, as seen in Figure~\ref{fig:Modes}.  In the case of slow conversions at both interfaces, one finds a continuous positive solution across the values of central densities of the stellar sequences, thus indicating that the stars are always stable, even on the descending branch of the mass--central-pressure~curve.

To summarize the recent findings regarding the stability of the hybrid stars, we have seen that
their stability against the fundamental oscillation modes strongly depends on the junction conditions at the interfaces between the phases.  These are determined by the rate of conversion between phases at the phase boundary. In the case of slow phase transitions (i.e., when the conversion time scale is larger than the characteristic period of the
oscillations), the usual stability criteria are modified and new stable segments appear that were previously unstable.  Alternative variants of junction conditions that are intermediate between slow and rapid conversion were also considered, but the resulting radial modes do not differ significantly from the slow conversion case, with corresponding implications for the stability of the stars~\cite{Rau2022}.

\section{Conclusions}
\label{sec:conclusions}

The investigation of dense QCD through the astrophysics of compact stars is an actively pursued
subject. This is due to the substantial observational progress, which includes measurements of the masses and radii of pulsars and gravitational wave signals from mergers of two neutron stars and neutron-star--black-hole binaries. A more thorough comprehension of the thermodynamics of dense QCD, weak interactions, and the dynamics of phase transitions would greatly enhance our ability to model astrophysical phenomena relevant to current observational programs.

This work gave an overview of the phase diagram of cold and dense QCD appropriate for compact stars.  We stressed that the universality of the phase diagram of imbalanced fermionic superfluids, such as cold atomic gases and nuclear matter, provides a valuable guide to the possible arrangement of the color-superconducting phases in neutron stars, the presence of tri-critical points, and BCS--BEC crossovers. The universality allows one to conjecture the possible structures of the phase diagram in the density--temperature plane including such  phases, such as the Fulde--Ferrel phase, deformed Fermi surface phase, and
the phase separation. 

As a novel contribution, the previously proposed parametrization of the EoS of dense quark matter with sequential phase transitions was extended to include a conformal fluid at large densities ($n\ge 10~n_{\rm sat}$) with the speed of sound $c_{\rm conf.}=1/\sqrt{3}$. The part of the $M$-$R$ diagram that contains twins and triplets remains intact because the transition to conformal fluid occurs at larger central densities than those achieved in these objects.  Nevertheless, for large central densities, we find behavior that is qualitatively different from earlier studies of this regime: the $M$--$R$ curves spiral in; i.e., after reaching a minimum, they turn to the right (larger radius region), thus avoiding the region of ultra-compact stars. Therefore, if the conformal limit is reached for densities much larger than those considered here, the ultracompact region with radii 6--7~km can be populated~\cite{Li2023PhRvD}. In the opposite case of the early onset of the conformal limit (as discussed in Section~\ref{sec:MR}), the radii will remain large, but small-mass regions can be populated if the stability criteria are modified by the slow conversion at the interface(s) between the phases. Another interesting new observation is that the change in the magnitude of the jump from 2SC to the CFL phase induces a special point on the $M$--$R$ diagram at which all the curves meet in analogy to the case of single-phase transition; see Ref.~\cite{Cierniak2020EPJST}.  The importance of studying this asymptotically large central density regime is phenomenologically relevant if the conversion between various quark and nuclear phases is slow compared to the characteristic timescale of oscillations, as discussed in Section~\ref{sec:modes}. In this case, the stars on the descending branch of mass--central-density (and its counterpart on the $M$-$R$ diagram) may be stable~\cite{Pereira2018,Curin2021,Goncalves2022,Rau2022},  contrary to the classical requirement $dM/d\rho_c >0$ for the branch to be stable, which in turn leads to
higher multipole (beyond triplets) stars on the $M$--$R$ diagram.

\vspace{6pt} \funding{This research was funded by Deutsche
  Forschungsgemeinschaft Grant No. SE 1836/5-2 and the Polish NCN
  Grant No. 2020/37/B/ST9/01937.  }

\dataavailability{The data presented in this study are available on
  request from the author.  }
 
 \acknowledgments{The author is grateful to M. Alford, J.-J. Li, and
   P. B. Rau for collaboration on modeling compact stars with quark
   cores and the referees for helpful comments.  }

\conflictsofinterest{The author declares no conflict of interest.}

\begin{adjustwidth}{-\extralength}{0cm}

\reftitle{References}

\PublishersNote{}
\end{adjustwidth}
\end{document}